\documentclass[prl,twocolumn,longbibliography,superscriptaddress]{revtex4-2}

\usepackage{amssymb}
\usepackage{amsmath}
\usepackage{amsfonts}
\usepackage{graphicx}
\usepackage{bm}
\usepackage{color}
\usepackage{multirow}
\usepackage{natbib}
\usepackage[normalem]{ulem}
\usepackage{relsize}
\usepackage{xspace}

\begin{document}
\title{Thermal transport across a Josephson junction in a dissipative environment}
\author{Tsuyoshi Yamamoto}
\affiliation{Faculty of Pure and Applied Physics, University of Tsukuba, Tsukuba, Ibaraki 305-8571, Japan}
\author{Leonid I. Glazman}
\affiliation{Department of Physics, Yale University, New Haven, Connecticut 06520, USA}
\author{Manuel Houzet}
\affiliation{Univ.~Grenoble Alpes, CEA, Grenoble INP, IRIG, PHELIQS, 38000 Grenoble, France}

\begin{abstract}
At zero temperature, a Josephson junction coupled to an ohmic environment displays a quantum phase transition between superconducting and insulating phases, depending whether the resistance of the environment is below or above the resistance quantum. At finite temperature, this so-called Schmid transition turns into a crossover. We determine the conditions under which the temperature dependence of the thermal conductance, which characterizes heat flow from a hot to cold resistor across the Josephson junction, displays universal scaling characteristic of the Schmid transition. We also discuss conditions for heat rectification to happen in the circuit. Our work can serve as a guide for identifying signatures of the Schmid transition in heat transport experiments.
\end{abstract}

\date{\today}

\date{\today}

\maketitle

{\it Introduction.--}
Heat transport mediated by microwave photons was observed in superconducting circuits operated at temperatures well below the superconducting transition. Ballistic heat transport characterized by the quantum of thermal conductance $G_q(T)=\pi k_B^2T/6\hbar$ at temperature $T$~\cite{Pendry1983} was shown up to distances of 50~$\mu$m~\cite{Timofeev2009} and 1~m~\cite{Partanen2016} in circuits with matched impedances between two resistors and a Josephson circuit connecting them. 

A practical device, a heat valve, relies on a controllable junction between two heat reservoirs. Flux-tunable heat valves were realized with superconducting circuits by connecting the reservoirs to a SQUID formed of a loop with two Josephson junctions, and applying a magnetic field through the loop~\cite{Meshke2006}; a heat valve controlled by gate was demonstrated with a Cooper pair transistor consisting of two junctions separated by a Coulomb island~\cite{Maillet2020}. The existing theory of heat propagation in such devices relies on perturbation theory in the strength of coupling realized by the controllable junction between the heat reservoirs. This limits theory applicability with respect to the range of circuit parameters and temperatures and excludes full consideration of nonperturbative effects, such as Schmid quantum phase transition~\cite{Schmid1983}.

Fortunately, a general theory of the quantum impurity problem allows one to overcome this limitation. The strategy is to leverage the knowledge of the impurity dynamical susceptibility for finding the thermal transport coefficients. A similar approach was used in finding the thermal conductance across a Kondo impurity~\cite{Saito2013} with the help of the dynamical susceptibility studied in~\cite{LeHur2012,Goldstein2013}.

\begin{figure}
\includegraphics[width=.9\columnwidth]{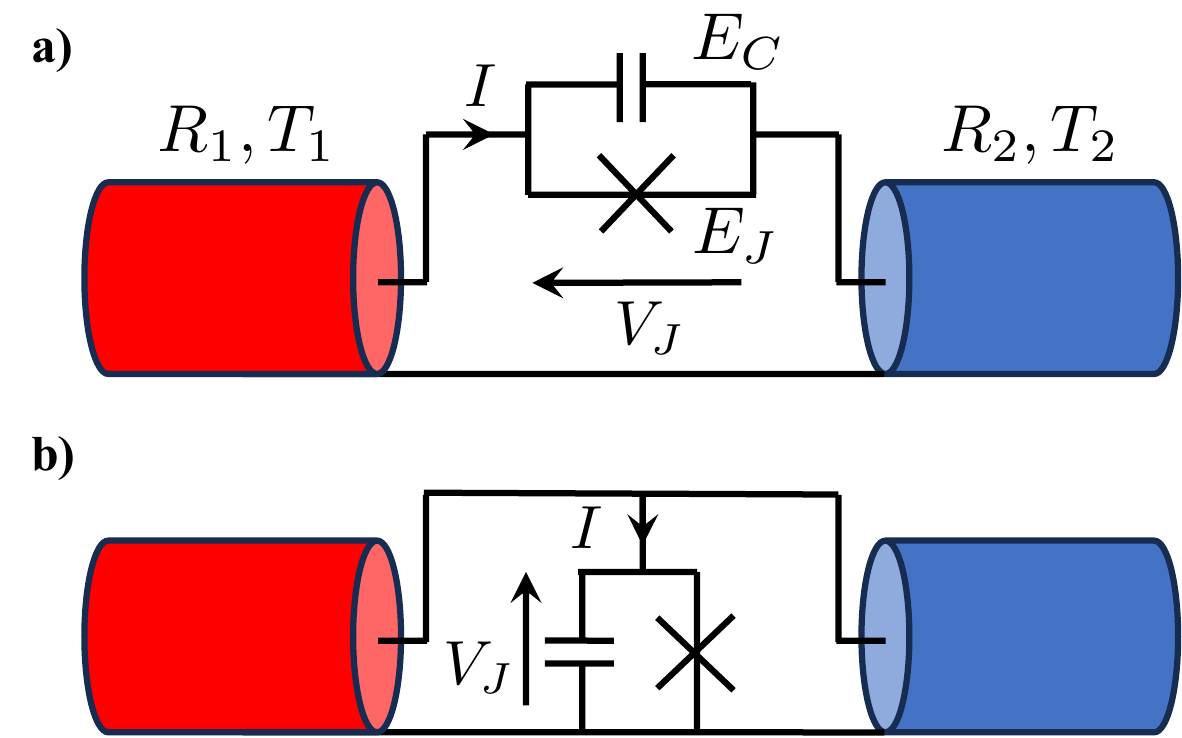}
\caption{\label{F:1} Two circuits for measuring heat transport. 
{\bf a)} Series configuration, {\bf b)} parallel configuration.}
\end{figure} 

In this work, we derive the relation between the thermal conductance $G_{\rm th}(T)$ and the admittance $Y(\omega, T)$ of the superconducting circuit at finite frequency $\omega$ and temperature $T$. After that, we extend the results for the admittance detailed in~\cite{Houzet2024} to finite temperatures and evaluate $G_{\rm th}(T)$ in the context of the Schmid transition between superconducting and insulating ground states, which is controlled by the dimensionless ratio of the resistance of the environment ``seen" by a Josephson junction to the resistance quantum, $R_Q=\pi \hbar/2e^2$.

The results are most clearly represented by the scaling function $g(t)$,
\begin{equation}
G_{\rm th}(T)=(4R_1R_2/R^2)G_q(T)g(t),
\end{equation} 
detailing the deviation from the ballistic transport prediction. Here $t=T/T_\star$ is the temperature normalized by the characteristic scale $T_\star$. The latter depends both on the Josephson junction parameters and a dimensionless parameter that characterizes the wave impedance seen by the junction, $K=R_Q/2R$ with $R=R_1+R_2$, where $R_1$ and $R_2$ are the baths' impedances in the circuit of Fig.~\ref{F:1}a.

\begin{figure}
\includegraphics[width=.9\columnwidth]{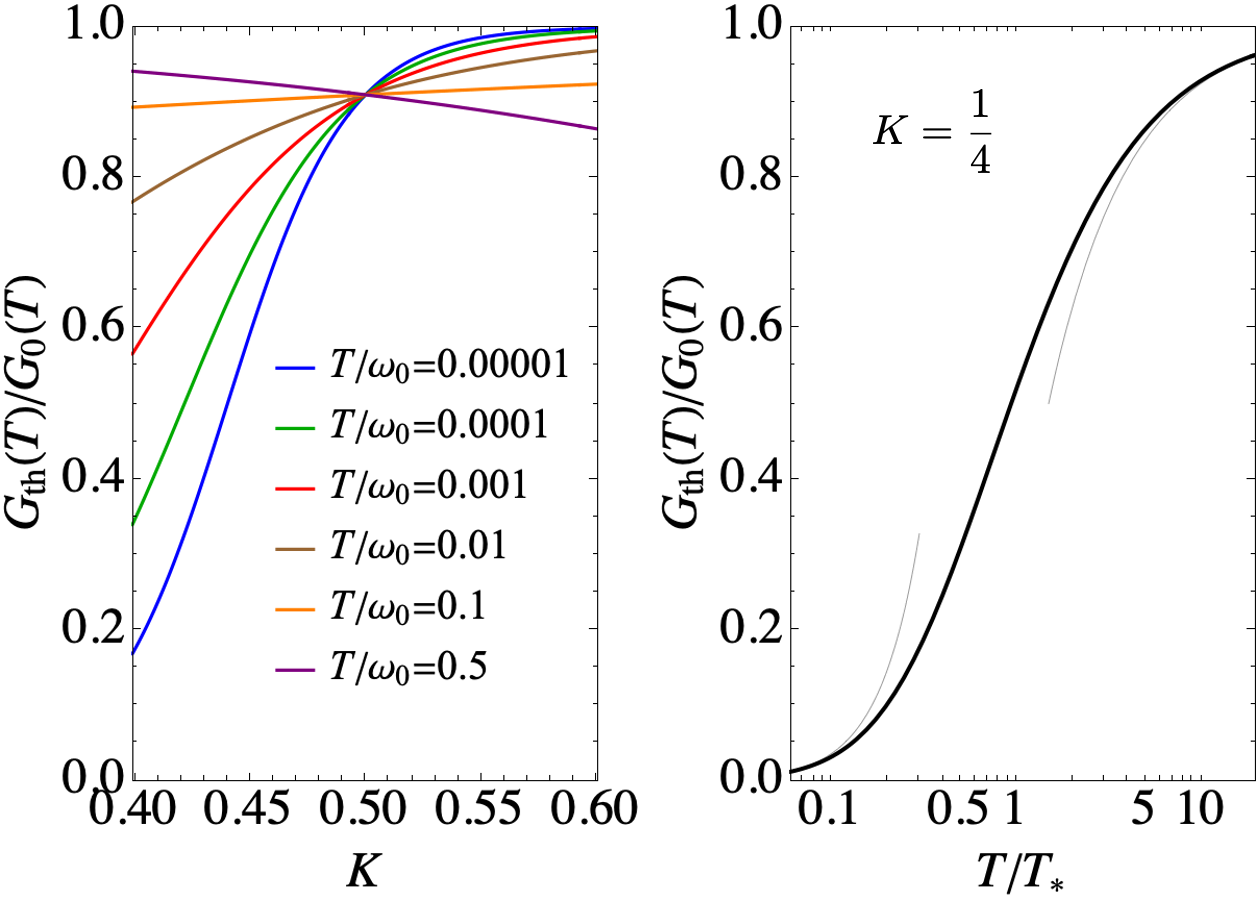}
\caption{\label{F:2} 
{\bf Left:} thermal conductance as a function of the environment impedance for a transmon with $\lambda/\omega_0=0.1$ at various temperatures. {\bf Right:} thermal conductance as a function of the temperature at the Toulouse point, $K=\frac14$. The thin lines are the low-$T$ and high-$T$ asymptotes.}
\end{figure}

The knowledge of the scaling function $g(t)$ allows us to find $G_{\rm th}(T)$ outside the domain of previously used perturbation theory. The newly found non-perturbative results include the heat conductance of a high-capacitance Josephson junction (a transmon) and a Cooper pair box (a charge qubit).
In a broader context, we relate the overall behavior of $g(t)$ to the Schmid transition: $g(t)$ is a monotonically increasing function of $t$ at $K<1/2$, its monotonicity is opposite at $K>1/2$,  see Fig.~\ref{F:2}. At the Schmid transition, $g(t)$ is temperature-independent. We find analytically the full scaling form of $g(t)$ in the vicinity of the transition at $K=1/2$ and at the Toulouse point ($K=1/4$) by mapping to a free fermion problem~\cite{Kane1992}.

The Schmid transition is currently attracting much attention both experimentally and theoretically~\cite{Houzet2020,Burshtein2021,Kuzmin2021,Leger2022,Kuzmin2023,Houzet2024,Burshtein2024,Kurilovich2024,Daviet2023,Masuki2022,Morel2021,Murani2020,Subero2023}. In particular, it was the focus of a recent heat transport experiment with highly dissipative resistors galvanically coupled to a flux-tunable SQUID~\cite{Subero2023}. We hope our work can contribute to the understanding of the measurements performed in that work.

{\it Relation between admittance and heat conductance.--}
Let us start with the formula for the thermal conductance in the series configuration of Fig.~\ref{F:1}a, where two resistances $R_1$ and $R_2$ are held at different temperatures $T$ and $T+\Delta T$ and connected by a Josephson junction. In linear response, $\Delta T\ll T$, the heat current from the hot to cold resistor is $P=G_{\rm th}\Delta T$, where $G_{\rm th}$ is the thermal conductance at temperature $T$. The latter can be related to the complex scattering phase $\delta(\omega,T)=\delta'(\omega,T)+i\delta''(\omega,T)$ off a circuit consisting of a Josephson junction in series with a resistor $R$, at finite frequency $\omega$ and temperature $T$. Equivalently, one may use the real part of the complex admittance of that circuit, 
$Y(\omega,T)=[1-e^{2i\delta(\omega,T)}]/2R$,
in order to express $G_{\rm th}$ as
\begin{equation}
\label{eq:Gth}
G_{\rm th}(T)=\frac{4R_1R_2}{R^2}\int_0^\infty \frac{d\omega}{2\pi}\frac {\omega^2/4T^2}{\sinh^2(\omega/2T)}\frac{1- \Re e^{2i\delta(\omega,T)}}2.
\end{equation}
Hereinafter we use units with $\hbar=k_B=1$.

To derive Eq.~\eqref{eq:Gth}, we may use boson scattering theory. For this, we describe the resistors that appear in Fig.~\ref{F:1} as transmission lines held at different temperatures $T_i$ ($i=1,2$).  We introduce incoming and outgoing bosonic modes at frequency $\omega$ in the lines, $A_i^{\rm in}(\omega)$ and $A_i^{\rm out}(\omega)$, such that the Fourier harmonics of the voltage and current at the line's end in contact with the junction are expressed as
\begin{subequations} 
\label{eq:VI}
\begin{eqnarray}
V_i(\omega)&=&\sqrt{R_i}\left[A_i^{\rm in}(\omega)+A_i^{\rm out}(\omega)\right],\\
I_i(\omega)&=&\frac1{\sqrt{R_i}}\left[A_i^{\rm in}(\omega)-A_i^{\rm out}(\omega)\right].
\end{eqnarray}
\end{subequations}
In the series configuration of Fig.~\ref{F:1}a, the current flowing through the junction is $I=I_1=-I_2$, while the voltage across the junction is $V_J=V_1-V_2$. We then define the elastic scattering matrix at frequency $\omega$, $S(\omega)=\{S_{ij}(\omega)\}$ ($i,j=1,2$), such that it relates the incoming and outgoing modes: $A_i^{\rm out}=S_{ij}A_j^{\rm in}$. Elimination of $V_i$ and $I_i$ in favor of $I$ and $V_J$ then allows us to express $S$ in a diagonalized form,
\begin{eqnarray}
\label{eq:S-diag}
S=U^T
{\rm diag}(e^{2i\delta},1)U,\quad
U=\frac1{\sqrt{R}}\left(\begin{array}{cc}\sqrt{R_1}&-\sqrt{R_2}\\ \sqrt{R_2}&\sqrt{R_1}\end{array}\right),
\end{eqnarray}
where $e^{2i\delta}=(V_J-RI)/(V_J+RI)$. The unitary matrix $U$ expresses that only one combination of the two lines' modes effectively couples to the junction. Thus, defining the admittance $Y\equiv I/V$ with $V=RI+V_J$, we recover the relation between $\delta$ and $Y$ given above Eq.~\eqref{eq:Gth}. Here let us emphasize that $\delta$ and $Y$ must be computed under the nonequilibrium conditions fixed by the different temperatures in the leads. For reservoirs connected to a purely reactive dipole, $\delta''=0$ and $S$ is unitary, such that boson scattering is purely elastic. In general, however, one should consider inelastic scattering in addition to the elastic cross section between two leads, $\sigma^{\rm el}_{12}(\omega)=(R_1R_2/R^2)|1-e^{2i\delta(\omega)}|^2$.

To address inelastic scattering, we then introduce the partial inelastic scattering cross section $\sigma_{j|i}(\omega'|\omega)$ for a boson with frequency $\omega$ in line $i$ to be converted into a boson with frequency $\omega'<\omega$ in line $j$. As the junction effectively couples to one combination of the two lines' modes only, see discussion below Eq.~\eqref{eq:S-diag}, we can relate $\sigma_{2|1}(\omega'|\omega)=\sigma_{1|2}(\omega'|\omega)=(R_1R_2/R^2)\sigma(\omega'|\omega)$ with the partial inelastic scattering cross section  $\sigma(\omega'|\omega)$ off a Josephson junction in series with a single transmission line with resistance $R$. Energy conservation imposes 
\begin{equation}
\int_0^\infty d\omega'\omega'\sigma(\omega'|\omega)=\omega \sigma^{\rm in}(\omega)
\end{equation}
with total inelastic cross section $\sigma^{\rm in}(\omega)=1-e^{-4\delta''(\omega)}$.

We may then use these relations to simplify the heat current between leads 1 and 2,
\begin{eqnarray}
\label{eq:heat-flux}
P&\equiv&\!\!\!\int\! \frac{d\omega}{2\pi}[n_1(\omega)-n_2(\omega)]\left\{\omega \sigma^{\rm el}_{12}(\omega)\!+\!\int \!d\omega'\sigma_{1|2}(\omega'|\omega)\right\}\nonumber\\
&=&\frac {4R_1R_2}{R^2}\int \frac{d\omega}{2\pi}\omega[n_1(\omega)-n_2(\omega)]\frac{1- \Re e^{2i\delta(\omega)}}2,
\end{eqnarray}
where $n_i(\omega)$ are Bose functions at temperatures $T_i$ and the admittance should be calculated for the athermal distribution $n(\omega)=[R_1n_1(\omega)+R_2n_2(\omega)]/R$. Taking $T=(T_1+T_2)/2$ and $\Delta T=T_1-T_2$, we readily recover Eq.~\eqref{eq:Gth} at $\Delta T\ll T$, where the admittance is now evaluated in equilibrium at temperature $T$. So far, nothing was assumed about the circuit connecting the two baths. For a linear circuit with pure elastic scattering, our results match those of~\cite{Pascal2011,Thomas2019}. Our formalism also allows recovering the many-body results of~\cite{Saito2013}. Reference~\cite{Saito2013} also established an equivalence of the many-body results with those obtained in~\cite{Segal2005} in the limit of weak coupling, but at arbitrary $\Delta T$.

Performing a similar calculation for the parallel configuration of Fig.~\ref{F:1}b, where two resistances $R_1$ and $R_2$ are connected to the same side of a junction that is grounded on its other side, yields the thermal conductance
\begin{equation}
\label{eq:para}
\tilde G_{\rm th}=G_0-G_{\rm th},\qquad G_0=({4R_1R_2}/{R^2})G_q.
\end{equation}
Here $G_0$ corresponds to Eq.~\eqref{eq:Gth} evaluated at $Y'(\omega)=1/R$; $G_0=G_q$ at matched impedances, $R_1=R_2$. Furthermore, the scattering phase that appears in Eq.~\eqref{eq:Gth} should be evaluated for a Josephson junction shunted by an impedance $R_1R_2/R$, i.e. $K=R_QR/2R_1R_2$. [To derive Eq.~\eqref{eq:para} we used different relations for the current through the junction, $I=I_1+I_2$, and the voltage, $V_J=V_1=V_2$.]

{\it Universal scaling.--}
Here we specify the results for the circuit of Fig.~\ref{F:1}a, where the junction has Josephson energy $E_J$ and charging energy $E_C=e^2/2C$ with capacitance $C$. We first focus on a transmon ($E_J\gg E_C$), suitable for demonstrating the scaling behavior on the insulating side of the transition ($K<\frac12$) in a broad range of temperatures $T\ll\omega_0$; here $\omega_0=\sqrt{8E_JE_C}$ is the Josephson plasma frequency. On the superconducting side of the Schmid transition, $K>\frac12$, a transmon behaves as an inductive short at all relevant temperatures, $G_{\rm th}(T \ll \omega_0)\approx G_0(T)$. 

At $K<\frac12$ in scaling regime, $G_{\rm th}(T)$ can be cast in the form
\begin{equation}
\label{eq:scaling}
G_{\rm th}(T)=G_0(T) g\left({T}/{T_\star},K\right),\qquad T\ll \omega_0,
\end{equation}
with crossover temperature 
\begin{equation}
\label{eq:crossover}
T_\star=\frac{\omega_0}{2\pi}\left(\sqrt{\frac{2K\Gamma^2(2K)}{\Gamma(4K)}}\frac{\pi\lambda}{\omega_0}\right)^{1/(1-2K)}.
\end{equation}
Here $\lambda=(8^5E_J^3E_C/\pi^2)^{1/4}e^{-\sqrt{8E_J/E_C}}$ is the phase slip amplitude. The capacitive-like response, $G_{\rm th}\to 0$ at $T\to 0$, indicates an insulating ground state. On the other hand, we expect $G_{\rm th}(T_\star \ll T\ll \omega_0)\approx G_0(T)$ when the junction is, again, essentially a short at the relevant frequencies.
Below we analyze the scaling function $g(t,K)$ and provide its low-$T$ and high-$T$ asymptotes. 

To address the asymptote at $T\gg T_\star$, we use a finite-temperature generalization~\cite{Aslangul1987} of Ref.~\cite{Houzet2024} to find the dissipative part of the admittance in a circuit with a transmission line terminated by a transmon,
\begin{equation}
\label{eq:G-high}
RY'(\omega)
=1-\left|\Gamma\left(2K+\frac{i\beta\omega}{2\pi}\right)\right|^2
\left(\frac{\beta\omega_\star}{2\pi}\right)^{2-4K}
\frac{\sinh(\beta\omega/2)}{\beta\omega/2}
\end{equation}
with $\beta=1/T$ and crossover frequency $\omega_\star=2\pi T_\star/[\Gamma(2K)]^{1/(1-2K)}$. Using Eqs.~\eqref{eq:Gth} and \eqref{eq:G-high}, we find the high-$T$ asymptote
\begin{equation}
\label{eq:high}
g(t\gg 1,K)=1-a_>(K)/t^{2-4K}
\end{equation}
with 
\begin{equation}
a_>(K)=\frac 3{\pi^2}\int_0^\infty dx\frac{x/2}{\sinh (x/2)}\frac{|\Gamma(2K+ix/2\pi)|^2}{\Gamma^2(2K)}.
\end{equation}
Note that at $K\to \frac12$ the $t$-dependence in Eq.~\eqref{eq:high} weakens, and $a_>(K)\to 1$.

At $T\ll T_\star$, we use a dual Hamiltonian valid at $\frac14<K<\frac12$~\cite{Houzet2024} to find
\begin{eqnarray}
RY'(\omega)
=
\tilde c(1/4K)\tilde c^{1/2K}(K)
\left|\Gamma\left(\frac1{2K}+\frac{i\beta\omega}{2\pi}\right)\right|^2
\nonumber\\
\times 
\left(\frac{\beta\omega_\star}{2\pi}\right)^{2- 1/K}\frac{\sinh(\beta\omega/2)}{\beta\omega/2}
\quad
\label{eq:G-low}
\end{eqnarray}
with $\tilde c(K)=8K^3\Gamma^2(2K)/\Gamma(4K)$. 
Inserting Eq.~\eqref{eq:G-low} into Eq.~\eqref{eq:Gth}, we find the low-$T$ asymptote of the scaling function,
\begin{equation}
\label{eq:above} 
g(t\ll 1,K)=a_<(K)t^{1/K-2}
\end{equation}
with 
\begin{equation}
a_<(K)=
 \frac {3b(K)}{\pi^2}\int_0^\infty dx\frac{x/2}{\sinh (x/2)}\frac{|\Gamma(1/2K+ix/2\pi)|^2}{\Gamma^2(1/2K)},
\end{equation}
and $b(K)=\tilde{\tilde c}(1/4K)\tilde{\tilde c}^{1/2K}(K)$, $\tilde{\tilde c}(K)=\tilde c(K)\Gamma^{2}(2K)$.
Here, like in Eq.~\eqref{eq:high}, the $t$-dependence weakens, and $a_<(K)\to 1$ on the approach of the critical point, $K\to \frac12$.

At $|K-\frac12|\ll 1$, both asymptotes \eqref{eq:G-high} and \eqref{eq:G-low} for the finite-frequency admittance are combined into a formula that describes the entire crossover at any $T,\omega\ll\omega_0$,
\begin{equation}
\label{eq:admittance-FFP}
RY'(\omega)=\left(1+\frac{\left|\Gamma\left(1+2\delta K+i\frac{\beta\omega}{2\pi}\right)\right|^2}{(\beta T_\star)^{4\delta K}\Gamma^2(1+2\delta K)}\frac{\sinh(\beta\omega/2)}{\beta\omega/2}\right)^{-1}
\end{equation}
with $\delta K=K-\frac12$. At the frequencies $\omega\sim T$ that are relevant for thermal transport, we may ignore the term $2\delta K$ in the argument of Gamma functions, such that Eq.~\eqref{eq:admittance-FFP} simplifies to $RY'(\omega\sim T)\approx1/ [1+(T/T_\star)^{4\delta K}]$. Inserting this result into Eq.~\eqref{eq:Gth} and using Eq.~\eqref{eq:crossover}, we find
\begin{equation}
\label{result}
G_{\rm th}(T)=\frac{G_0(T)}{1+{\cal T}(2\pi T/\omega_0)^{4\delta K}},\quad {\cal T}=\left(\frac{\pi \lambda}{\omega_0}\right)^2\ll 1
\end{equation}
In the left panel of Fig.~\ref{F:2}, we use Eq.~\eqref{result} to plot the $K$ dependence of the heat conductance at various fixed temperatures, as $K$ varies across the transition point. We observe that the normalized heat conductance increases/decreases with $T$ on the insulating/superconducting side of the Schmid transition. As expected, $G_{\rm th}$ remains close to $G_0$ at $K>\frac 12$, while a full crossover from 0 to $G_0$ can be observed at $K<\frac12$. At $K=\frac 12$, the ratio $G_{\rm th}(T)/G_0(T)$ is $T$-independent.

At $K<\frac14$, Eq.~\eqref{eq:above} is not applicable as it would predicted a stronger suppression-in-$T$ of heat conductance than the one of a capacitor, $G_{\rm th}\propto T^3$. In fact, the correct answer originates from a capacitive contribution to $Y(\omega)$, not captured by the dual Hamiltonian~\cite{Guinea1995,Houzet2024},
\begin{eqnarray}
\label{eq:admittance-lowK}
RY'(\omega)&=&(\omega/\beta(K)\omega_\star)^2,\\
\frac1{\beta(K)}\!\!\!&=&\!\!\!\frac1{2\sqrt{\pi}}
{\Gamma\left(\frac{1/2}{1-2K}\right)\Gamma\left(\frac{1-3K}{1-2K}\right)}
\left(\frac{\tilde c(K)}{4K^2}\right)^{\frac1{2(1-2K)}}.
\nonumber
\end{eqnarray}
It dominates over the contribution \eqref{eq:G-low} and does not depend on $T$. Inserting Eq.~\eqref{eq:admittance-lowK} into \eqref{eq:Gth} we find
\begin{equation}
\label{eq:lowT2}
g(t\ll 1,K)=\frac{4 \pi^2}5\frac {[\Gamma(2K)]^{2/(1-2K)}}{(2\pi \beta( K))^2} t^2.
\end{equation}

Note that both Eqs.~\eqref{eq:above} and \eqref{eq:lowT2} yield the same $T^3$-dependence for $G_{\rm th}(T)$ at $K\to\frac14\pm0$ with different prefactors. This indicates a non-analytical dependence of $G_{\rm th}$ around the Toulouse point $K=\frac14$.  At that point, we may use the exact free-fermion solution~\cite{Kane1992} to find
\begin{equation}
\label{eq:Y-Tououse}
RY(\omega)=1+\frac {2\omega_\star}{i\pi \omega}\left[\psi\left(\frac 12+\frac{2\omega_\star-i\pi \omega}{2\pi^2 T}\right)-\psi\left(\frac 12+\frac{\omega_\star}{\pi^2 T}\right)\right],
\end{equation}
and
\begin{eqnarray}
\label{eq:Toulouse}
g(t,K=\frac14)=
1+\frac 3{\pi^3 t}\int_0^\infty dx\frac{x}{\sinh^2(x/2)}\times\qquad\qquad\\
\qquad\qquad\times{\Im}\left[\psi\left(\frac12 +\frac{2}{\pi^2 t}-\frac{i x}{2\pi}\right)
       -\psi\left(\frac12 +\frac{2}{\pi^2 t}\right)\right].\nonumber
\end{eqnarray}
Equation~\eqref{eq:Toulouse} reproduces the high-$T$ asymptote $g(t\gg1)=1-3/4t$, Eq.~\eqref{eq:high} at $K=\frac14$; it also reproduces the low-$T$ asymptote given by the sum of Eqs.~\eqref{eq:above} and \eqref{eq:lowT2}, $g(t\ll 1,K=\frac14)=3\pi^4t^2/80$. The result is illustrated in the right panel of Fig.~\ref{F:2}. The relative contributions of Eqs.~\eqref{eq:above} and \eqref{eq:lowT2} to the result at $K=\frac14$ were obtained in \cite{Kane1996}, where a similar issue was studied in the context of an impurity in a Luttinger liquid. Here we also find the absolute amplitude of the effect in terms of the circuit parameters.

At $K\to 0$, the transmon is almost disconnected and behaves as a capacitor with capacitance $C_\star=e^2/\pi^2\lambda$~\cite{Houzet2024}, such that $Y^{-1}(\omega)=R+i/\omega C_\star$ at $T\ll T_\star=\lambda/\sqrt{2}$. Inserting the admittance into Eq.~\eqref{eq:Gth} we find that the crossover in $G_{\rm th}$ from 0 to $G_0$ actually occurs on the temperature scale $ K T_\star\ll T_\star$, with asymptotes $g(t\ll K)=t^2/40K^2$, in agreement with Eq.~\eqref{eq:lowT2} at $K\to 0$, and $g(K\ll t\ll 1)=1-3\sqrt{8}K/t$. At higher temperature, $T\sim T_\star$, the already small correction to $g= 1$ changes to decay even faster, $g(t\gg 1)=1-6K/t^2$, cf.~Eq.~\eqref{eq:high} at $K\to 0$.

{\it Duality.--} By duality~\cite{Schmid1983}, for a charge qubit ($E_J\ll E_C$) on the superconducting side of the Schmid transition, $K >\frac12$,
\begin{equation}
\label{eq:scaling-qb}
G_{\rm th}(T)=G_0\left[1- g\left({T}/{\Theta_\star},1/{4K}\right)\right], \qquad T\ll \Gamma,
\end{equation}
with the same function $g(t)$ as in the transmon case studied above, and with another crossover temperature $\Theta_\star$ obtained from $T_\star$ of Eq.~\eqref{eq:crossover} after substitutions: $\lambda\to E_J$, $\omega_0\to 2e^{\gamma}\Gamma$ ($\gamma\approx0.58$ is Euler's constant), 
and $K\to 1/4K$, with plasma resonance linewidth $2\Gamma=1/RC$. The ground state at $K>\frac12$ is superconducting, leading to inductive-like response,
$G_{\rm th}(T\to 0)= G_0$.

{\it Rectification.--}
Phase $\delta$ in Eq.~\eqref{eq:heat-flux} depends on the distribution $n(\omega)=[R_1n_1(\omega)+R_2n_2(\omega)]/R$ and, thus, on temperatures of both baths.
This naturally leads to heat rectification~\cite{Segal2005}, a difference in heat currents at opposite signs of the temperature bias, provided the device is asymmetric ($R_1\neq R_2)$. Heat rectification was recently measured in a superconducting circuit~\cite{Senior2020}. To quantify this effect in the circuit of Fig.~\ref{F:1}a connecting a hot bath at temperature $T$ and cold one at $T=0$, we assume $R_1\gg R_2$. In this case $n(\omega)$ is an equilibrium distribution at temperature $T$ or at $T=0$, depending on the sign of the temperature bias. These assumptions allow us characterizing heat rectification with the ratio 
\begin{equation}
\label{eq:ratio}
{\cal R}=\frac{\int_0^\infty d\omega\omega Y'(\omega,T)/(e^{\omega/T}-1)}{\int_0^\infty d\omega\omega Y'(\omega,0) /(e^{\omega/T}-1)},
\end{equation}
where we used the relation between $\delta$ and $Y$. Using Eq.~\eqref{eq:G-low} at $\frac14<K<\frac12$ and $T\ll T_\star$, we find that both numerator and denominator in Eq.~\eqref{eq:ratio} have the same temperature dependence $\propto  T^{1/K}$, but different prefactors. Thus $\cal R$ only depends on $K$, and we find that is decreases monotonically from 11 to 1 as $K$ increases from $\frac14$ to $\frac12$. On the other hand, we find that the rectification effect vanishes (${\cal R}=1$) at $T\gg T_\star$, as well as at $T\ll T_\star$ and $K<\frac14$, where $Y$ is essentially $T$-independent at relevant $\omega$, see Eqs.~\eqref {eq:G-high} and \eqref{eq:admittance-lowK}, respectively. Inserting Eq.~\eqref{eq:Y-Tououse} into \eqref{eq:ratio} at the Toulouse point, we also find that ${\cal R}$ decreases monotonically from 7/2 to 1 as $T$ increases.

{\it Discussion.--}
In this work we used the scattering theory for interacting bosons to find a compact formula that relates heat conductance to the finite-frequency admittance of a Josephson junction in series with resistors at different temperatures. We analyzed the emergent scaling behavior of the thermal response of the circuit, and elucidated the manifestation of the Schmid transition in the thermal conductance as a crossing point at $K=\frac 12$ of the scaled thermal conductances measured at different temperatures. We hope our work provides a guide for identifying clear signatures that would confirm the existence of the Schmid transition in future heat transport experiments.

A recent heat transport experiment performed with a flux-tunable SQUID challenged the existence of the transition because the thermal conductance was still depending on the flux, despite the circuit being on the insulating side of the transition~\cite{Subero2023}. According to our study, there is no contradiction. Indeed, the characteristic temperature scale, which enters the response of a transmon, clearly depends on the flux through the dependence of the phase slip amplitude on the flux-tunable Josephson energy. In our opinion, recent experiments on dual Shapiro steps~\cite{Shaikhaidarov2022,Crescini2023,Kaap2024,Kaap2024b}, which can only be understood within the paradigm of the Schmid transition, leave little doubt on the existence of the insulating phase. Nevertheless understanding quantitatively the finite-frequency~\cite{Murani2020} or thermal~\cite{Subero2023} response in recent experiments remains challenging.

{\it Note added.--} Recently we learned about a study~\cite{ALY2024} of the charge qubit limit on the insulating side of the transition, which are complementary to the results presented in this work.

\begin{acknowledgments}
We thank A. Levy Yeyati for sending us the manuscript of~\cite{ALY2024} prior to making it public, as well as T. Kato, A. Levy Yeyati, O. Maillet, and J. Pekola for illuminating discussions. This work was supported by NSF Grant No. DMR-2002275, ARO Grant No. W911NF-22-1-0053, the French CNRS IRP HYNATOQ and ANR-23-CE47-0004. TY acknowledges support from JST Moonshot R\&D–MILLENNIA Program (Grant No. JPMJMS2061). 
\end{acknowledgments}


\begin{thebibliography}{50}

\bibitem{Pendry1983}
J. B. Pendry, J. Phys. A {\bf 16}, 2161 (1983).

\bibitem{Timofeev2009}
A. V. Timofeev, M. Helle, M. Meschke, M. M\"{o}tt\"{o}nen, and J. P. Pekola, 
Phys. Rev. Lett.~{\bf 102}, 200801 (2009).

\bibitem{Partanen2016}
M. Partanen, K. Y. Tan, J. Govenius, R. E. Lake, M. K. M\"{a}kel\"{a}, T. Tanttu, and M. M\"{o}tt\"{o}nen, Nature Physics {\bf 12}, 460 (2016).


\bibitem{Meshke2006}
M. Meschke, W. Guichard, and J. P. Pekola, Nature {\bf 444}, 187 (2006).

\bibitem{Maillet2020}
O. Maillet, D. Subero, J. T. Peltonen, D. S. Golubev, and J. P. Pekola, Nat. Commun. {\bf 11}, 4326 (2020).

\bibitem{Schmid1983}
A. Schmid, Phys. Rev. Lett. {\bf 51}, 1506 (1983).

\bibitem{Saito2013}
K. Saito and T. Kato, Phys. Rev. Lett. {\bf 111}, 214301 (2013).

\bibitem{LeHur2012}
K. Le Hur
Phys. Rev. B {\bf 85}, 140506(R) (2012).

\bibitem{Goldstein2013}
M. Goldstein, M. H. Devoret, M. Houzet, and L. I.  Glazman, Phys. Rev. Lett. {\bf 110}, 017002 (2013).

\bibitem{Houzet2024}
M. Houzet, T.  Yamamoto, and L. I. Glazman, arXiv:2308.16072 (2023).




\bibitem{Kane1992}
C. L. Kane and M. P. A. Fisher,
Phys. Rev. B {\bf 46}, 15233 (1992).

\bibitem{Subero2023}
D. Subero, O. Maillet, D. S. Golubev, G. Thomas, J. T. Peltonen, B. Karimi, M. Mar\'in-Su\'arez, A. Levy Yeyati, R. S\'anchez, S. Park, and J. P. Pekola, Nature Communications {\bf 14}, 7924 (2023).



\bibitem{Murani2020}
A. Murani, N. Bourlet, H. le Sueur, F. Portier, C. Altimiras, D. Esteve, H. Grabert, J. Stockburger, J. Ankerhold, and P. Joyez, Phys. Rev. X {\bf 10}, 021003 (2020).

\bibitem{Houzet2020}
M. Houzet and L. I. Glazman,
Phys. Rev. Lett. {\bf 125}, 267701 (2020).

\bibitem{Burshtein2021}
A. Burshtein, R. Kuzmin, V. E. Manucharyan, and M. Goldstein,
Phys. Rev. Lett. {\bf 126}, 137701 (2021).

\bibitem{Kuzmin2021}
R. Kuzmin, N. Grabon, N. Mehta, A. Burshtein, M. Goldstein, M. Houzet, L. I. Glazman, and V. E. Manucharyan,
Phys. Rev. Lett. {\bf 126}, 197701 (2021).

\bibitem{Morel2021}
T. Morel and C. Mora,
Phys. Rev. B {\bf 104}, 245417 (2021).

\bibitem{Masuki2022}
K. Masuki, H. Sudo, M. Oshikawa, and Y. Ashida,
Phys. Rev. Lett. {\bf 129}, 087001 (2022); 
T. S\'{e}pulcre, S. Florens, and I. Snyman,
Phys. Rev. Lett. {\bf 131}, 199701 (2023);
K. Masuki, H. Sudo, M. Oshikawa, and Y. Ashida,
Phys. Rev. Lett. {\bf 131}, 199702 (2023).


\bibitem{Daviet2023}
R. Daviet and N. Dupuis,
Phys. Rev. B {\bf 108}, 184514 (2023).

\bibitem{Leger2022}
S. L\'eger, T. S\'epulcre, D. Fraudet, O. Buisson, C. Naud, W. Hasch-Guichard, S. Florens, I. Snyman, D. M. Basko, and N. Roch,
SciPost Phys. {\bf 14}, 130 (2023).

\bibitem{Kuzmin2023}
R. Kuzmin, N. Mehta, N. Grabon, R. A. Mencia, A. Burshtein, M. Goldstein, and V. E. Manucharyan,
arXiv:2304.05806v1.


\bibitem{Burshtein2024} 
A. Burshtein and M. Goldstein, arXiv:2308.15542.

\bibitem{Kurilovich2024}
V. D. Kurilovich, B. Remez, and L I. Glazman,
arXiv:2403.04624.

\bibitem{Pascal2011}
L. M. A. Pascal, H. Courtois, and F. W. J. Hekking, Phys. Rev. B {\bf 83}, 125113 (2011).

\bibitem{Thomas2019}
G. Thomas, J. P. Pekola, and D. S. Golubev, 
Phys. Rev. B {\bf 100}, 094508 (2019).

\bibitem{Segal2005}
D. Segal and A. Nitzan, Phys. Rev. Lett. {\bf 94}, 034301 (2005).



\bibitem{Aslangul1987}
C. Aslangul, N. Pottier, D. Saint-James,
Journal de Physique, {\bf 48}, 1093 (1987).





\bibitem{Guinea1995}
F. Guinea, G. G\'{o}mez Santos, M. Sassetti, and M. Ueda,
EPL {\bf 30}, 561 (1995).



\bibitem{Kane1996}
C. L. Kane and M. P. A. Fisher,
Phys. Rev. Lett. {\bf 76}, 3192 (1996).



\bibitem{Senior2020}
J. Senior, A. Gubaydullin, B. Karimi, J. T. Peltonen, J. Ankerhold, and J. P. Pekola, Communications Physics {\bf 3}, 40 (2020).



\bibitem{Shaikhaidarov2022}
R. S. Shaikhaidarov, K. H. Kim, J. W. Dunstan, I. V. Antonov, S. Linzen, M. Ziegler, D. S. Golubev, V. N. Antonov, E. V. Il’ichev, and O. V. Astafiev, 
Nature {\bf 608}, 45 (2022).
\bibitem{Crescini2023}
N. Crescini, S. Cailleaux, W. Guichard, C. Naud, O. Buisson, K. W. Murch, and N. Roch, 
Nature Physics {\bf 19}, 851 (2023).
\bibitem{Kaap2024}
F. Kaap, C. Kissling, V. Gaydamachenko, L. Gr\"{u}nhaupt, and S. Lotkhov, 
arXiv:2401.06599 [cond-mat.mes-hall].
\bibitem{Kaap2024b}
F. Kaap, D. Scheer, F. Hassler, and S. Lotkhov, 
Phys. Rev. Lett. {\bf 132}, 027001 (2024).
\bibitem{ALY2024}
A. Levy Yeyati, D. Subero, J. P. Pekola, and R. S\'anchez (2024).




\end{thebibliography}
\end{document}